\documentclass[draftclsnofoot,onecolumn,twoside,12pt]{IEEEtran}
\usepackage{graphicx}
\usepackage{bbm}

\usepackage{amsfonts,amssymb,amsmath}

\def\RR{{\mathbb R}}

\def\EE{{\mathbb E}}

\def\Ex#1{{\EE\left\{#1\right\}}}

\def\CC{\mathbb C}

\def\NN{\mathbb N}

\def\dd{\;\mathrm{d}}

\def\ii{\mathrm{i}}

\def\be{\begin{equation}}
\def\beq#1{\begin{equation}\label{#1}}
\def\ee{\end{equation}}
\def\bea{\begin{eqnarray}}
\def\beqa#1{\begin{eqnarray}\label{#1}}
\def\eea{\end{eqnarray}}
\def\ba{\begin{array}}
\def\ea{\end{array}}

\def\defeq{\stackrel{\Delta}{=}}

\DeclareMathAlphabet{\mathpzc}{OT1}{pzc}{m}{it}

\def\ccA{{\mathscr A}}

\def\ccD{{\mathscr D}}

\def\ccL{{\mathscr L}}

\def\ccM{{\mathscr M}}

\def\cO{{\mathcal O}}

\def\ccS{{\mathscr S}}

\def\cV{{\mathcal V}}
\def\cW{{\mathcal W}}
\def\tcW{\widetilde{\mathcal W}}

\def\bs{{\mathbf s}}

\def\ba{{\mathbf a}}

\def\bLambda{\boldsymbol{\Lambda}}
\def\bw{{\mathbf w}}
\def\bW{{\mathbf W}}

\def\bC{{\mathbf C}}

\def\bI{{\mathbf I}}

\def\btau{\boldsymbol{\tau}}
\def\btheta{\boldsymbol{\theta}}
\def\bTheta{\boldsymbol{\Theta}}

\usepackage{type1cm,eso-pic,color}

\usepackage{algorithm,algorithmic}

\usepackage{mathrsfs} 
\usepackage{mathpazo}
\usepackage[scaled=.95]{helvet}
\usepackage{courier}
\usepackage{xspace}

\newtheorem{theorem}{Theorem}
\newtheorem{proposition}{Proposition}

\newtheorem{remark}{Remark}
\graphicspath{{figures/}}

\newcommand{\eg}{e.g.\@\xspace}
\newcommand{\ie}{i.e.\@\xspace}

\def\beq{\begin{equation}}
\def\eeq{\end{equation}}

\allowdisplaybreaks[4]

\begin{document}
\bstctlcite{MyBSTcontrol}

\title{Spectral analysis for nonstationary audio}
\author{Adrien~Meynard and~Bruno~Torr\'esani%
\thanks{A. Meynard and B. Torr\'esani are with Aix Marseille Univ, CNRS, Centrale Marseille, I2M, Marseille, France.}%
%\thanks{Part of this work was done when the two authors were at Centre de Recherches Math\'ematiques, UMI 3457, CNRS and Universit\'e de Montr\'eal, Canada.}%
%\thanks{This paper has supplementary downloadable material available at http://ieeexplore.ieee.org., provided by the author. The material includes a technical report containing proofs of mathematical statements and additional examples. Contact bruno.torresani@univ-amu.fr for further questions about this work}
%\thanks{Manuscript received ???, 2017; revised ???, 2018.}
}
\date{\today}
%\markboth{IEEE/ACM Transactions on Audio, Speech, and Language Processing,~Vol.~??, No.~??, Month~20??}%
%{Shell \MakeLowercase{\textit{et al.}}: Spectral analysis for nonstationary audio}

\maketitle

\begin{abstract}
A new approach for the analysis of nonstationary signals is proposed, with a focus on audio applications. Following earlier contributions, nonstationarity is modeled via stationarity-breaking operators acting on Gaussian stationary random signals. The focus is on time warping and amplitude modulation, and an approximate maximum-likelihood approach based on suitable approximations in the wavelet transform domain is developed. This paper provides theoretical analysis of the approximations, and introduces JEFAS, a corresponding estimation algorithm. The latter is tested and validated on synthetic as well as real audio signal.
\end{abstract}

% Note that keywords are not normally used for peerreview papers.
\begin{IEEEkeywords}
Nonstationary signals, deformation, wavelet analysis, time warping, local spectrum, Doppler effect
\end{IEEEkeywords}
%\vspace{-1mm}
\section{Introduction}
\label{se:intro}

Nonstationarity is a key feature of acoustic signals, in particular audio signals. For example, a large part of information carried by musical and speech signals is encoded by their nonstationary nature. This is also the case for environment sounds (for example, nonstationarity of car noises or wind informs about speed variations), and many animals (\eg bats and dolphins) use nonstationary signals for localization and communication. Beyond acoustics, amplitude and frequency modulation are of prime importance in many domains such as telecommunication.

While stationarity can be given rigorous definitions, nonstationarity is a very wide concept, as there are infinitely many ways to depart from stationarity. The theory of random signals and processes (see~\cite{Koopmans95spectral,Priestley82spectral} and references therein) gives a clear meaning to the notion of stationarity. In the context of time series analysis, Priestley~\cite{Priestley82spectral,Priestley88nonlinear} was one of the first to develop a systematic theory of nonstationary processes, introducing the class of locally stationary processes and the notion of evolutionary spectrum. A similar approach was followed in~\cite{Mallat98adaptive}, who proposed a wavelet-based approach to covariance estimation for locally stationary processes (see also~\cite{Nason00wavelet}). An alternate theory of locally stationary time series was developed by Dahlhaus~\cite{Dahlhaus00likelihood} (see also~\cite{VonSachs00wavelet} for a corresponding stationarity test). In a different context, frequency-modulated stationary signal were considered in~\cite{Wisdom14extending,Omer13estimation}, and time warping models were analyzed in~\cite{Wisdom14Improved}. In several of these approaches, wavelet, time-frequency and similar representations happen to play a key role for the characterization of nonstationarity.

In a deterministic setting, a popular nonstationarity model expresses the signal as a sum of $K$ sinusoidal components  $y(t) = \sum_{k=1}^K A_k(t) \cos(2\pi\phi_k(t))$. This model has been largely used in speech processing since early works by McAulay and Quatieri~\cite{McAulay86speech} (see~\cite{Lagrange07enhancing} and references therein for more recent developments, and~\cite{Depalle93analysis,Godsill02bayesian} for probabilistic approaches). %For instance, this model can be used to represent the echolocation signal emitted by a bat.
The instantaneous frequencies $\phi_k'$ of each mode give important information about the physical phenomenon. Under smoothness assumptions on functions $A_k$ and $\phi'_k$, techniques such as ridge/multiridge detection (see~\cite{Carmona98practical} and references therein), synchrosqueezing or reassignment have been developed to extract theses quantities from a single signal observation (see~\cite{Auger13time,Wu13instantaneous,Lin18wave} for recent accounts).

In sound processing, signals often possess a harmonic structure, which corresponds to a special case of the above model where each instantaneous frequency $\phi_k'$ is multiple of a fundamental frequency $\phi_0'$: $\phi_k'(t) = (k+1)\phi_0'(t)$.  In the special case $A_k(t) = \alpha_k A_0(t)$, we can describe such signals as a stationary signal $x(t) = \sum_{k=1}^K \alpha_k \cos(2\pi k t + \varphi_k) $ modified by time warping and amplitude modulation: $y(t) = A_0(t)x(\phi_0(t))$.
A major limit of this model is that each component is purely sinusoidal while audio signals often contain broadband information. However, sounds originating from physical phenomena can often be modeled as stationary signals, deformed by a stationarity-breaking operator (\eg time warping, amplitude modulation). For example, sounds generated by a variable-speed engine or any stationary sound deformed by Doppler effect can be described as such. A stochastic time warping model has been introduced in~\cite{Clerc02texture,Clerc03estimating}, where wavelet-based approximation and estimation techniques were developed. In~\cite{Omer13estimation,Omer17time}, an approximate maximum-likelihood approach was proposed for the joint estimation of the time warping and power spectrum of the underlying Gaussian stationary signal, exploiting similar approximations.

In this paper, we build on results of~\cite{Omer13estimation,Omer17time} which we extend and improve in several ways. We develop an approximate maximum likelihood method for estimating jointly time warping and amplitude modulation (not present in~\cite{Omer13estimation,Omer17time}) from a single realization. While the overall structure of the algorithm is similar, we formulate the problem as a continuous parameter estimation problem, which avoids quantization effects present in the earlier approaches, and allows computing a Cram\'er-Rao bound for assessing the precision of the estimate. After completing the estimation, the inverse deformation can be applied to the input signal, which yields an estimate for the power spectrum. 

The outline of the paper is as follows. After giving some definitions and notations in Section~\ref{se:notations}, we detail in Section~\ref{se:estimation} the nonstationary signal models we consider, and specify the assumptions made on underlying stationary signals. We also analyze the effect of time warping and amplitude modulation in the wavelet domain, which we exploit in designing the estimation procedure. We finally propose an alternate estimation algorithm and analyze the expected performances of the corresponding estimator. Section~\ref{se:results} is devoted to numerical results on synthetic and real signals. We also shortly describe in this section an extension published in~\cite{Meynard17spectral} involving simultaneously time warping and frequency modulation. Mathematical developments are given as supplementary material, together with additional examples.

\section{Notations and background}
\label{se:notations}
\subsection{Random signals and stationarity}
\label{sub:stationarity}
Throughout this paper, we will work in the framework of the theory of random signals. Signals of interest will be modeled as realizations of random processes $t\in\RR\to X_t\in\CC$. Signals of interest are real-valued, however we will work with complex-valued functions since complex-valued wavelet transforms will be used. In this paper, the random processes will be denoted by uppercase letters while their realizations will be denoted by lowercase letters.  The random processes will be assumed to have null mean $(\Ex{X_t=0}$ for all $t$) and be second-order, \ie they have a well-defined covariance kernel $\Ex{X_t\overline{X}_{t'}}$. A particularly interesting class of such stochastic processes is the class of second order (or weakly) stationary processes, for which $C_X(t-t')\defeq\Ex{X_t\overline{X}_{t'}}$ is a function of $t-t'$ only. Under these assumptions, the Wiener-Khinchin theorem states that the covariance kernel may be expressed as the inverse Fourier transform of a nonnegative measure $d\eta_X$, which we will assume to be continuous with respect to the Lebesgue measure: $\dd\eta_X(\nu)=\ccS_X(\nu)\dd\nu$, for some nonnegative $L^1$ function $\ccS_X$ called the power spectrum. We then write
\[
C_X(t)=\int\ccS_X(\nu)e^{2\ii\pi\nu t}\dd\nu\ .
\]
We refer to textbooks such as~\cite{Koopmans95spectral,Priestley82spectral} for a more complete mathematical account of the theory, and to~\cite{Omer17time} for an extension to the setting of distribution theory.

%\vspace{-2mm}
\subsection{Elementary operators}
\label{sub:operators}
Our approach rests on nonstationary models obtained by deformations of stationary random signals. We will mainly use as elementary operators the 
pointwise multiplication $A_\alpha$, translation $T_\tau$, dilation $D_s$, and frequency modulation $M_\nu$ defined as follows:
\begin{align*}
A_\alpha x(t) &= \alpha x(t)\,, & T_\tau x(t) &= x(t - \tau)\,,\\
D_s x(t) &= q^{-s/2}x(q^{-s}t)\,, & M_\nu x(t) &= e^{2\ii\pi\nu t}x(t)\,.
\end{align*}
where $\alpha,\tau,s,\nu\in\RR$ and $q>1$ is a fixed number.

The amplitude modulation commutes with the other three operators, which satisfy the commutation rules
%\begin{align*}
%T_\tau D_s &= D_s T_{q^{-s}\tau}\,, \\
%T_\tau M_\nu &= e^{-2i\pi\nu\tau}M_\nu T_\tau\,, \\
%M_\nu D_s &= D_s M_{\nu q^s}\,.
%\end{align*}
\[
T_\tau D_s \!=\! D_s T_{q^{-s}\tau}\,, \ 
T_\tau M_\nu \!=\! e^{-2\ii\pi\nu\tau}M_\nu T_\tau\,, \ 
M_\nu D_s \!=\! D_s M_{\nu q^s}\,.
\]
\subsection{Wavelet transform}
\label{sub:wavelets}
Our analysis relies heavily on transforms such as the continuous wavelet transform (and discretized versions). In particular,
the wavelet transform of a signal $X:t\in\RR\to X_t$ is defined as:
\be
\label{fo:WT}
\cW_X(s,\tau) \defeq \langle X,\psi_{s\tau} \rangle\,,\ \hbox{with}\ 
\psi_{s\tau} = T_\tau D_s\psi\,.
\ee
where $\psi$ is the analysis wavelet, \ie a smooth function with fast decay away from the origin. It may be shown that, for suitable choices of $\psi$, the wavelet transform is invertible (see~\cite{Carmona98practical}), but we will not use that property here. Notice that realizations of a continuous time random process generally do not decay at infinity. However, for a suitably smooth and localized wavelet $\psi$, the wavelet transform can still be well defined (see~\cite{Carmona98practical,Omer17time} for more details). In such a situation the wavelet transform of $X$ is a two-dimensional random field, which we analyze in the next section. Besides, in this paper the analysis wavelet $\psi$ is complex-valued and belongs to the space $H^2(\RR) = \left\lbrace \psi\in L^2(\RR)\ :\ \mbox{supp}(\hat\psi)\subset\RR^+ \right\rbrace$. In that framework, a useful property is that, if $X$ is a real, zero-mean, Gaussian process, then $\cW_X$ is a complex, zero-mean, circular, Gaussian random field.

Classical choices of wavelets in $ H^2(\RR)$ are (analytic) derivative of Gaussian $\psi_k$ (which has $k$ vanishing moments), and the sharp wavelet $\psi_\sharp$ (with infinitely many vanishing moments) introduced in~\cite{Meynard17spectral}. These can be defined in the positive Fourier domain by
\begin{equation}
\label{fo:wavelet}
\hat\psi_k(\nu) = \nu^k e^{-k\nu^2/2\nu_0^2}\, ,\quad  \hat\psi_\sharp(\nu) = \epsilon^{\frac{\delta(\nu,\nu_0)}{\delta(\nu_1,\nu_0)}}\, ,\quad \nu>0
\end{equation}
and vanish on the negative Fourier half axis. Here $\nu_0$ is the mode of $\hat\psi$ and $\nu_1$ is a cutoff frequency. $\epsilon$ is a prescribed numerical tolerance chosen so that $\hat\psi_\sharp(\nu_1)=\epsilon$, and the divergence $\delta$ is defined by $\delta(a,b) = \frac1{2}\left(\frac{a}{b} +\frac{b}{a}\right)-1$. The quality factor of $\psi_\sharp$, \ie the  center frequency to bandwidth ratio, can be expressed as $Q = 1/\sqrt{C(C+4)}$, where $C=-\delta(\nu_1,\nu_0)\ln 2/\ln{\epsilon}\ >0$ (see supplementary material for more details).

\begin{remark}
In~\eqref{fo:WT}, the scale constant $q>1$ acts a unit selector  for the scale $s$. For example, in musical terminology, $q=2$ means that $s$ is measured in octaves, whereas for $q=2^{1/12}$, $s$ is measured in semitones. 
\end{remark}

\subsection{Amplitude modulation, time warping}
The nonstationary signals under consideration are obtained as linear deformations of stationary random signals. 
Deformations of interest here are amplitude modulations and time warpings. Amplitude modulations are pointwise multiplications by smooth functions,% defined as 
\beq
\label{fo:amp.modul}
\ccA_a:\qquad \ccA_a x(t) = a(t) x(t)\ ,
\ee
where $a\in C^1$ is a real valued function, such that
\beq
\label{fo:am.control}
0<c_a \le a(t) \le C_a<\infty,\qquad \forall t\ ,
\ee
for some constants $c_a, C_a\in\RR_+^*$. Time warpings are compositions with smooth and monotonic functions,
\beq
\label{fo:time.warp}
\ccD_\gamma:\qquad \ccD_\gamma x(t) = \sqrt{\gamma'(t)} x(\gamma(t))\ ,
\ee
where $\gamma\in C^2$ is a strictly increasing smooth function, satisfying the control condition~\cite{Omer17time}
\beq
\label{fo:time.warp.control}
0<c_\gamma \le \gamma'(t) \le C_\gamma<\infty,\qquad \forall t\ ,
\ee
for some constants $c_\gamma, C_\gamma\in\RR_+^*$.

Amplitude modulations constitute a simple model for nonstationarity. While there exists a well established state of the art for demodulation algorithms in deterministic settings (in particular in telecommunications), the stochastic case has attracted less attention. In the recent literature, one may mention~\cite{Wisdom14Improved}, where the so-called DEMON spectrum is proposed for amplitude modulation estimation. This problem will be tackled here using wavelet transform.

Time warping is an important transformation which has been exploited in various contexts, starting from Doppler effect, which we briefly address at the end of this paper, but also speech processing, bioacoustics (see~\cite{Stowell18computational} and references therein) and more generally in diverse fields such as chemistry or bioinformatics.
The reference algorithm for time warping estimation is DWT (Dynamic Time Warping, see~\cite{Senin08DTW} for a review), which has been successfully applied to speech processing, in particular speech recognition. However, DTW is essentially a template matching algorithm, and does not address the problem considered here, where no template is available.
Closer to our point of view are approaches based upon transforms such as the Harmonic transform~\cite{Zhang04harmonic} or the Fan-Chirp transform~\cite{Weruaga07fanchirp} (see~\cite{Dunn07sinewave} for an application to speech analysis/synthesis). These mainly involve computing a time-frequency representation of a warped copy of the input signal. Even though this does not seem to be strictly necessary, the warping function generally belongs to a parametrized family (for example quadratic), and has to be estimated. The application domain seems to be limited so far to locally harmonic, deterministic signal models, and the estimation of the warping function strongly relies on these assumptions.
Wavelet transform and scalogram are also natural representations for estimating warping. Actually, interpreting (normalized) time slices of the signal scalogram as probability distributions naturally suggests to compute a time-dependent average scale, directly related to the value of the warping function. This is the approach we will use as baseline approach. However, our approximations below allow us to give a more precise meaning to this remark; in addition, scalogram lacks the phase information which turns out to be quite relevant and yield more precise estimations.

The approach we develop below exploits complex valued wavelet transform, and combines amplitude modulation and time warping in the framework of a generic stochastic signal model, without any harmonicity assumption. This allows us to set the corresponding estimation problems as statistical inference problems, and use tools from estimation theory.

\section{Joint estimation of time warping and amplitude modulation}
\label{se:estimation}

\subsection{Model and approximations}
Let us first describe the deformation model we will mainly be using in the following. 

Assume one is given a (unique) realization of a random signal of the form
\begin{equation}
\label{eq:model}
Y = \ccA_a\ccD_\gamma X
\end{equation}
where $X$ is a stationary zero-mean real random process with (unknown) power spectrum $\ccS_X$.
The goal is to estimate the deformation functions $a$ and $\gamma$ from this realization of $Y$, exploiting the stationarity of $X$.
%The goal is to estimate jointly the deformation functions $a$ and $\gamma$ and the power spectrum $\ccS_X$ from this realization of $Y$, exploiting the assumed stationarity of $X$.
\begin{remark}
\label{rem:identifiability}
The stationarity assumption is not sufficient to yield unambiguous estimates, as affine functions $\gamma(t)=\lambda t + \mu$ do not break stationarity: for any stationary $X$, $\ccD_\gamma X$ is stationary too. Therefore, the warping function $\gamma$ can only be estimated up to an affine function, as analyzed in~\cite{Clerc03estimating,Omer17time}. Similarly, the amplitude function $a$ can only be estimated up to a constant factor.
\end{remark}

Key ingredients here are the smoothness of the functions $a$ and $\gamma$, and their slow variations. This allows us to perform a local analysis using smooth and localized test functions, on which the action of $\ccA_a$ and $\ccD_\gamma$ can be approximated by their so-called \textit{tangent operators} $\widetilde{\ccA_a^\tau}$ and $\widetilde{\ccD_\gamma^\tau}$ (see~\cite{Clerc03estimating,Omer13estimation,Omer17time,Omer15Modeles}). Given a test function $g$ located near $t=\tau$ (\ie decaying fast enough as a function of $|t-\tau|$), Taylor expansions near $t=\tau$ yield
\begin{align}
\label{fo:approx.am}
\ccA_a g(t)&\approx\widetilde{\ccA_a^\tau} g(t)\,,\  \hbox{with}\ \widetilde{\ccA}_a^\tau \defeq A_{a(\tau)}\ ,\\
\label{fo:approx.warping}
\ccD_\gamma g(t)&\approx\widetilde{\ccD_\gamma^\tau} g(t)\,,\  \hbox{with}\ \widetilde{\ccD_\gamma^\tau} \defeq T_\tau D_{\!-\log_q\!(\!\gamma'(\tau)\!)}T_{\!-\gamma(\tau)}\, .
\end{align}
Therefore, we approximate the wavelet transform of $Y$ by
$
\cW_Y(s,\tau)\approx \widetilde{\cW}_Y(s,\tau)\defeq\left\langle \widetilde{\ccA_a^\tau}\widetilde{\ccD_\gamma^\tau} X,T_\tau D_s\psi\right\rangle$, \ie
\be
\label{fo:approx.wavelet.transform}
\widetilde{\cW}_Y(s,\tau)=a(\tau)\cW_X\left(s+\log_q(\gamma'(\tau)),\gamma(\tau)\right)\ .
\ee
Here, we have used the standard commutation rules of translation and dilation operators given in Section~\ref{sub:operators}.

The result below provides a quantitative assessment of the quality of the approximation. Hereafter, we denote by $\|f\|_\infty=\mathsf {ess\, sup}_t |f(t)|$ the essential absolute supremum of a function $f$.
\smallskip
\begin{theorem}
\label{th:approx}
Let $X$ be a second order zero-mean stationary random process, let $Y$ be the nonstationary process defined in~\eqref{eq:model}. Let $\psi$ be a smooth test function, localized in such a way that $|\psi (t)|\le 1/(1\!+\!|t|^\beta)$ for some $\beta>2$. Let $\cW_Y$ be the wavelet transform of $Y$, $\tcW_Y$ its approximation given in~\eqref{fo:approx.wavelet.transform}, and let $\varepsilon=\cW_Y-\tcW_Y$ denote the approximation error. Assume $\psi$ and $\ccS_X$ are such that
\[
I_X^{(\rho)} \defeq \sqrt{\int_0^\infty\xi^{2\rho}\ccS_X(\xi)\dd\xi} <\infty\,,\  \hbox{where}\ \rho = \frac{\beta-1}{\beta+2}\ .
\]
Then the approximation error $\varepsilon$ is a second order, two-dimensional complex random field, and
\begin{equation*}
\Ex{\!|\varepsilon(s,\!\tau)|^2\!}\!\le\! C_a^2 q^{3s}\!\!\left(\!K_1\!\|\gamma''\|_\infty \!+\!
K_2 q^{\mu s}\|\gamma''\|_\infty^\rho\!+\!K_3\!\left\|a'\right\|_\infty\right)^{\! 2}
%\sqrt{C_\gamma\int\left(\xi\|\gamma''\|_\infty + \|\alpha''\|_\infty\right)^{2\frac{\lambda-1}{\lambda+2}}\ccS_X(\xi)d\xi}
\end{equation*}
where 
\begin{align*}
K_1 &= \frac{\beta\sigma_X}{2(\beta-2)\sqrt{c_\gamma}}\ ,& K_2 &= I_X^{(\rho)}\left(\frac{\pi}{2}\right)^{\rho}\!\sqrt{C_\gamma}\,\frac{4}{3\rho}\ ,\\
K_3 &= \dfrac{\sqrt{C_\gamma}\beta\sigma_X}{(\beta-2)c_a}\ , 
&\mu &= \frac{\beta-4}{\beta+2}\ ,
\end{align*}
$\sigma_X^2$ being the variance of $X$.
\end{theorem}
\smallskip

The proof, which is an extension of the one given in~\cite{Omer17time}, rests on Taylor approximations of $\gamma'$ and $a$ in the neighborhood of $t=\tau$ and subsequent integral majorizations, is given as supplementary material.

\begin{remark}
The assumption on $\beta$ ensures that the parameters belong to the following intervals: $1/4 <\rho <1$ and $-1/2 <\mu <1$. Therefore, the variance of the approximation error tends to zero when the scales are small (\ie $s\to-\infty$).
Besides, the error is inversely proportional to the speed of variations of $\gamma'$ and $a$. This is consistent with the approximations of the deformation operators by their tangent operators made in equations~\eqref{fo:approx.am} and~\eqref{fo:approx.warping}.
\end{remark}

From now on, we will assume the above approximations are valid, and work on the approximate random fields. The problem is then to estimate jointly $a$, $\gamma$ from $\tcW_Y$, which is a zero-mean random field with covariance
\be
\label{fo:approx.wavelet.covariance}
\Ex{\widetilde\cW_Y(s,\tau)\overline{\widetilde\cW_Y(s',\tau')}} = C(s,s',\tau,\tau')
\ee
where the kernel $C$ reads
\begin{align}
\label{fo:covariance}
\nonumber
C(s,s'\!,\tau,&\tau')\!= \;a(\tau)a(\tau')
q^{\frac{s+s'}{2}}\sqrt{\gamma'(\tau)\gamma'(\tau')}\int_0^\infty \ccS_X(\xi) \\
 \times &\overline{\hat{\psi}}\!\left(q^{s}\gamma'\!(\tau)\xi\right)
\hat{\psi}\!\left(\!q^{s'}\!\gamma'\!(\tau')\xi\!\right)e^{2\ii\pi\xi(\gamma(\!\tau\!)-\gamma(\!\tau'\!)\!)}\!\dd\xi.
\end{align}

%\vspace{-7mm}
\subsection{Estimation}

\subsubsection{Estimation procedure}
Our goal is to estimate both deformation functions $\gamma$ and $a$ from  the approximated wavelet transform $\tcW_y$ of a realization $y$ of $Y$, assuming the latter is a reliable approximation of the true wavelet transform. From now on, we additionally assume that $X$ is a Gaussian random process. Therefore, $\tcW_Y$ is a zero-mean circular Gaussian random field and its probability density function is characterized by the covariance matrix. However, equation~\eqref{fo:covariance} shows that besides deformation functions the covariance also depends on the power spectrum $\ccS_X$ of the underlying stationary signal $X$, which is unknown too. Therefore, the evaluation of the maximum likelihood estimate for $a$ and $\gamma$ requires a guess for $\ccS_X$. This constraint naturally brings the estimation strategy to an alternate algorithm. In~\cite{Meynard17spectral}, an estimate for $\ccS_X$ was obtained at each iteration using a Welch periodogram on a ``stationarized" signal $\ccA_{\tilde a}^{-1}\ccD_{\tilde\gamma}^{-1}Y$, $\tilde a$ and $\tilde\gamma$ being the current estimates for the deformation functions $a$ and $\gamma$. We use here a simpler estimate, computed directly from the wavelet coefficients.
The two steps of the estimation algorithm are detailed below.
\begin{remark}
The alternate likelihood maximization strategy is reminiscent of the Expec\-tation-Maximization (EM) algorithm, the power spectrum being the nuisance parameter. However, while it would be desirable to apply directly the EM paradigm (whose convergence is proven) to our problem, the dimensionality of the latter (and the corresponding size of covariance matrices) forces us to make additional simplifications that depart from the EM scheme. Therefore we turn to a simpler approach with several dimension reduction steps.
\end{remark}

\medskip
\textit{(a)\ Deformation estimation.} Assume that the power spectrum $\ccS_X$ is known (in fact, only an estimate $\tilde\ccS_X$ is known). Thus, we are able to write the likelihood corresponding to the observations of the wavelet coefficients. Then the maximum likelihood estimator is implemented to determine the unknown functions $\gamma$ and $a$.

The wavelet transform~\eqref{fo:WT} is computed on a regular time-scale grid $\bLambda=\bs\times\btau$, $\delta_s$ being the scale sampling step and $F_s$ the time sampling frequency. The sizes of $\bs$ and $\btau$ are respectively denoted by $M_s$ and $N_\tau$.

Considering the covariance expression~\eqref{fo:covariance} we want to estimate the vector of parameters $\bTheta = (\btheta_1, \btheta_2,\btheta_3) \defeq (a(\btau)^2,\log_q\left(\gamma'(\btau)\right), \gamma(\btau))$.
Let $\bW_y=\tcW_y(\bLambda)$ denote the discretized transform and let $\bC_{\bW}(\bTheta)$ be the corresponding covariance matrix. The related log-likelihood is
\be
\label{fo:approx.log.likelihood.2}
\ccL(\bTheta) = -\dfrac{1}{2}\ln\left|\det(\bC_\bW(\bTheta))\right| - \dfrac{1}{2}\bC_\bW(\bTheta)^{-1}\bW_y\cdot\bW_y\ .
\ee
The matrix $\bC_\bW(\Theta)$ is a matrix of size $M_sN_\tau\times M_sN_\tau$, which is generally huge. For instance, for a 5 seconds long signal, sampled at frequency $F_s=44.1$ kHz, when the wavelet transform is computed on 8 scales, the matrix $\bC_\bW(\bTheta)$ has about 3.1 trillion elements which makes it numerically intractable. In addition, due to the redundancy of the wavelet transform, $\bC_\bW(\bTheta)$ turns out to be singular, and likelihood evaluation is impossible.

To overcome these issues, we use a block-diagonal regularization of the covariance matrix, obtained by forcing to zeros entries corresponding to different time indices. In other words, we disregard time correlations in the wavelet domain, which amounts to considering fixed time vector $\bw_{y,\tau_n}=\tcW_y(\bs,\tau_n)$ as independent circular Gaussian vectors with zero-mean and covariance matrix
\be
\bC(\Theta_n)_{ij}=\theta_{n,1} \bC_{0}(\theta_{n,2})_{ij}\,,\ 1\leq i,j\leq M_s\,,
\ee
where
\begin{equation}
\label{fo:new.covariance}
\bC_{0}(\theta_{n,2})_{ij} = q^{(s_i+s_j)/2}\int_0^\infty \ccS_X(q^{-\theta_{n,2}}\xi)\overline{\hat{\psi}}(q^{s_i}\xi)\hat{\psi}(q^{s_j}\xi)\dd\xi\,.
\end{equation}
In this situation, the regularized log likelihood $\ccL^r$ splits into a sum of independent terms
\[
\ccL^r(\bTheta) = \sum_n \ccL(\Theta_n)\ ,
\]
where $\Theta_n = (\theta_{n,1},\theta_{n,2}) \defeq (\btheta_{1}(n),\btheta_{2}(n))$ corresponds to the amplitude and warping parameters at fixed time $\tau_n = \btau(n)$. Notice that, in such a formalism, $\theta_{n,3}=\gamma(\tau_n)$ does not appear anymore in the covariance expression.
Thus, we are led to maximize independently for each $n$
\be
\label{fo:approx.log.likelihood.time}
\ccL(\Theta_n) = -\frac{1}{2}\!\ln\left|\det(\bC(\Theta_n))\right| - \frac{1}{2}\bC(\Theta_n)^{-1}\bw_{y,\tau_n}\cdot\bw_{y,\tau_n}\ .
\ee
%To determine the maximum likelihood estimator $\widetilde\Theta_n$ of $\Theta_n$, $\ccL(\Theta_n)$ is maximized with respect to each component of $\Theta_n$.
For simplicity, the estimation procedure is done by an iterative algorithm (given in more details in part~\ref{ssse:algo}), which rests on two main steps. First, the log-likelihood is maximized with respect to $\theta_{n,2}$ using a gradient ascent method, for a fixed value of $\theta_{n,1}$. Second, for a fixed $\theta_{n,2}$, an estimate for  $\theta_{n,1}$ is directly obtained which reads
\begin{equation}
\label{fo:am.estim}
\tilde\theta_{n,1} = \dfrac{1}{M_s}\bC^{-1}_{0}(\theta_{n,2})\bw_{y,\tau_n}\cdot\bw_{y,\tau_n}\,.
\end{equation}

\medskip
\textit{(b)\ Spectrum estimation.}
Assume the amplitude modulation and time-warping parameters $\btheta_1$ and $\btheta_2$ are known (in fact, only estimates $\tilde\btheta_1$ and $\tilde\btheta_2$ are known). For any $n$ we can compute the wavelet transform 
\begin{equation}
\label{fo:wavelet.stat}
\frac{1}{\theta_{n,1}^{1/2}}\tcW_y\left(s-\theta_{n,2},\tau_n\right) = \cW_x(s,\gamma\left(\tau_n)\right)\,, 
\end{equation} 
For fixed scale $s_m$, $\bw_{x,s_m} \defeq \cW_x(s_m,\gamma(\btau))\in\CC^{N_\tau}$ is a zero-mean random circular Gaussian vector with time independent variance (as a realization of the wavelet transform of a stationary process). Hence, the empirical variance is an unbiased estimator of the variance. 
%Normalizing it by $\|\psi\|_{2}^2$, 
We then obtain the so-called wavelet spectrum
\begin{eqnarray}
\label{fo:spectrum.estimation}
\ccS_{X,\psi}(q^{-s_m}\omega_0)\!\!&\!\defeq\!&\!
\EE\left\lbrace\!\dfrac{1}{N_\tau\|\psi\|_{2}^2}\|\bw_{x,s_m}\|^2\!\right\rbrace\\
\!&\!=\!& \!\!\dfrac{1}{\|\psi\|_{2}^2}\!\int_0^\infty \!\!\!\!\!\ccS_X(\xi)q^{s_m}\!\left|\hat\psi\left(q^{s_m}\xi\right)\right|^2\! \dd\xi \, ,\qquad
%\nonumber
\end{eqnarray}
where $\omega_0$ is the central frequency of $|\hat\psi|^2$. $\ccS_{X,\psi}$ is a narrowband version of $\ccS_X$ centered around frequency $\nu_m = q^{-s_m}\omega_0$. Besides, the bandwidth of the filter is proportional to the frequency $\nu_m$. This motivates the introduction of the following estimator $\tilde\ccS_X$ of $\ccS_X$
\begin{equation}
\label{fo:spectrum.estimator}
\tilde\ccS_X(q^{-s_m}\omega_0) \defeq \dfrac{1}{N_\tau\|\psi\|_{2}^2}\|\bw_{x,s_m}\|^2\,.
\end{equation}
Finally, the estimate $\tilde{\ccS}_X$ is extended to all $\xi\in[0,F_s/2]$ by linear interpolation.

\subsubsection{Algorithm}
\label{ssse:algo}
The estimation procedure is implemented in an iterative alternate optimization algorithm. This algorithm whose pseudo-code is given as Algorithm~\ref{alg:estimation} is named \textit{Joint Estimation of Frequency, Amplitude, and Spectrum} (JEFAS). The initialization needs an initial guess for the power spectrum $\ccS_X$ of $X$. We use the spectrum estimator~\eqref{fo:spectrum.estimator} applied to the observation $Y$.

After $k$ iterations of the algorithm, estimates $\tilde\Theta_{n}^{(k)}$ and $\tilde\ccS_X^{(k)}$ for $\Theta_{n}$  and $\ccS_X$ are available. Hence we can only evaluate the plug-in estimate $\tilde\bC_0^{(k)}$ of $\bC_0$, obtained by replacing the power spectrum with its estimate in the covariance matrix~\eqref{fo:new.covariance}. This yields an approximate expression $\ccL^{(k)}$ for the log-likelihood, which is used in place of $\ccL$ in~\eqref{fo:approx.log.likelihood.time} for maximum likelihood estimation. The influence of such approximations on the performances of the algorithm are discussed in section~\ref{sse:performances}.

To assess the convergence of the algorithm, the relative update of the parameters is chosen as stopping criterion:
%\begin{equation}
%\label{fo:stop.crit}
%\dfrac{\left\Vert \tilde\btheta^{(k)}_j-\tilde\btheta^{(k-1)}_j \right\Vert_2^2}{\left\Vert \tilde\btheta^{(k-1)}_j \right\Vert_2^2} < \Lambda\,,\ \mbox{for}\  j=1,2\ ,
%\end{equation}
\begin{equation}
\label{fo:stop.crit}
\left\Vert \tilde\btheta^{(k)}_j-\tilde\btheta^{(k-1)}_j \right\Vert_2^2\Bigg/\left\Vert \tilde\btheta^{(k-1)}_j \right\Vert_2^2 < \Lambda\,,\ \mbox{for}\  j=1,2\ ,
\end{equation}
where $0<\Lambda<1$ is a user defined threshold.

Finally, after convergence of the algorithm to the estimated value $\tilde\bTheta^{(k)}$, $\log_q(\gamma')$ and $a^2$ are estimated through time by cubic spline interpolation. Besides, $\gamma$ is given by numerical integration assuming that $\gamma(0)=0$.

\begin{algorithm}
\caption{JEFAS (Joint Estimation of Frequency, Amplitude and Spectrum)}
\label{alg:estimation}
\begin{algorithmic}

\STATE {\bf Initialization:} Compute an estimate $\tilde\ccS_Y$ of the power spectrum of $Y$ as an initial guess $\tilde\ccS_X^{(0)}$ for $\ccS_X$. Initialize the estimator of the squared amplitude modulation with $\tilde \theta_{n,1}^{(0)}=1,\,\forall n$.

\STATE Compute the wavelet transform $\bW_y$ of $y$.

\STATE $k:=1$

\WHILE{ criterion~\eqref{fo:stop.crit} is false and $k\le k_{max}$}

\STATE $\bullet$ For each $n$, subsample $\bw_{y,\tau_n}$ on scales $\bs_p$, and estimate $\tilde\theta_{n,2}^{(k+1)}$ by maximizing the approximate log-likelihood $\ccL^{(k)}\left(\tilde\theta_{n,1}^{(k)},\theta_{n,2}\right)$ in~\eqref{fo:approx.log.likelihood.time}.
% with respect to $\theta_{n,2}$  for  .

\STATE $\bullet$ For each $n$, estimate $\tilde\theta_{n,1}^{(k+1)}$ by maximizing the approximate log-likelihood $\ccL^{(k)}\left(\theta_{n,1},\tilde\theta_{n,1}^{(k+1)}\right)$ with respect to $\theta_{n,1}$ in~\eqref{fo:approx.log.likelihood.time}. Or, in absence of noise, directly apply equation~\eqref{fo:am.estim}  using the regularized covariance matrix given by~\eqref{fo:regularization}.

\STATE $\bullet$ Construct the estimated wavelet transform $\bW_x$ of the underlying stationary signal by interpolation from $\bW_y$ and $\tilde\btheta^{(k)}$ with equation~\eqref{fo:wavelet.stat}. Estimate the corresponding power spectrum $\tilde\ccS_X^{(k+1)}$ with~\eqref{fo:spectrum.estimator}.

\STATE $\bullet$ $k:= k+1$

\ENDWHILE

\STATE $\bullet$ Compute $\tilde a$ and $\tilde\gamma$ by interpolation from $\tilde\bTheta^{(k)}$.
\end{algorithmic}
\end{algorithm}

\begin{remark}
To control the variances of the estimators, and the computational cost, two different discretizations of the scale axis are used for $\tilde{\btheta}_1$ or $\tilde{\btheta}_2$. Indeed, the computation of the log-likelihood involves the evaluation of the inverse covariance matrix. In~\cite{Omer17time}, a sufficient condition for invertibility was given in the presence of noise. The major consequence induced by this condition is that when $\delta_s$ is close to zero (\ie the sampling period of scales is small), the covariance matrix could not be numerically invertible. 
The scale discretization must then be sufficiently coarse to ensure good conditioning for the matrix. While this condition can be reasonably fulfilled to estimate $\theta_{n,2}$ without impairing the performances of the estimator, it cannot be applied to the estimation of $\theta_{n,1}$ because of the influence of $M_s$ on its Cram\'er-Rao bound (see section~\ref{sse:performances} below). The choice we made is to maximize $\ccL(\Theta_n)$ for $\theta_{n,2}$ with $\bw_{y,\tau_n}$ corresponding to a coarse sampling $\bs_p$ which is a subsampled version of the original vector $\bs$, the scale sampling step and the size of $\bs_p$ being respectively $p\delta_s$ and $\lfloor M_s/p\rfloor$ for some $p\in\NN^*$. While $\ccL(\Theta_n)$ is maximized for $\theta_{n,1}$ on the original fine sampling $\bs$, a regularization of the covariance matrix has to be done to ensure invertibility.
%The regularized matrix is $\bC_{0,r}(\theta_{n,2})$ where the spectrum expression $\ccS_X$ is replaced by $\ccS_{Z,r}$ defined as
The regularized matrix is constructed by replacing covariance matrix $\bC_0(\theta_{n,2})$ in~\eqref{fo:new.covariance} by its regularized version $\bC_{0,r}(\theta_{n,2})$, given by
\begin{equation}
\label{fo:regularization}
\bC_{0,r}(\theta_{n,2}) = (1-r)\bC_{0,r}(\theta_{n,2}) + r\bI\ ,
\end{equation}
for some regularization parameter $0 \leq r \leq 1$.
\end{remark}

\begin{remark}
After convergence of the estimation algorithm, the estimated functions $\tilde a$ and $\tilde\gamma$ allow constructing a ``stationarized" signal
$$
\tilde x = \ccD_{\tilde\gamma^{-1}}\ccA_{\tilde a^{-1}}y\ .
$$
$\tilde x$ is an estimation of the original underlying stationary signal $x$. Furthermore, the Welch periodogram~\cite{Welch67use} may be computed from $\tilde x$ to obtain an estimator of $\ccS_X$ whose bias does not depend on frequency (unlike the estimator used within the iterative algorithm).
\end{remark}

\begin{remark}
In order to accelerate the speed of the algorithm, the estimation can be done only on a subsampled time grid. The main effect of this choice on the algorithm concerns the final estimation of $a$ and $\gamma$ which is more sensitive to the interpolation operation.
%the estimation of the power spectrum because as $N_\tau$ is smaller the estimator bias can potentially improve as shown by equation~\eqref{fo:spectrum.bias}.
\end{remark}

In the following section, we analyze quantities that enable the evaluation of the expected performances of the estimators, and their influence on the algorithm. The reader who is not directly interested
in the statistical background may skip this section and jump directly to the numerical results in part~\ref{se:results}. 

\subsection{Performances of the estimators and the algorithm}
\label{sse:performances}
%\medskip
\textit{(a)\ Bias.}
For $\theta_{n,1}$, the estimator is unbiased when the actual values of $\theta_{n,2}$ and $\ccS_X$ are known. In our case, the bias $b_{n,1}^{(k)}\left(\theta_{n,1}\right)=\Ex{\tilde\theta_{n,1}^{(k)}}-\theta_{n,1}$ is written as
\begin{equation}
\label{fo:bias.am}
b_{n,1}^{(k)}\left(\theta_{n,1}\right)\! = \!\dfrac{\theta_{n,1}}{M_s}\ \mbox{Trace}\left\lbrace \tilde\bC_0^{(k)}\!\!\left(\tilde\theta_{n,2}^{(k)}\right)^{-1}\bC_0(\theta_{n,2}) \!-\! \bI \right\rbrace\,.
\end{equation}
As expected, the better the covariance matrix estimation, the lower the bias $\left|b_{n,1}^{(k)}\right|$.

For $\theta_{n,2}$, as we do not have a closed-form expression for the estimator we are not able to give an expression of the bias. Nevertheless, if we assume that the two other true variables are known, as a maximum likelihood estimator we make sure that $\tilde\theta_{n,2}$ is asymptotically unbiased (\ie $\tilde\theta_{n,2} \to \theta_{n,2}$ when $M_s\to\infty$).

Regarding $\ccS_X$, equation~\eqref{fo:spectrum.estimation} shows that the estimator yields a smoothed, thus biased version of the spectrum. Proposition~\ref{pr:spectrum.bias} below shows that the estimated spectrum converges to this biased version when the deformation parameters converge to their actual values.
\vskip1mm
\begin{proposition}
\label{pr:spectrum.bias}
Let $\psi\in H^2(\RR)$ be an analytic wavelet such that $\hat\psi$ is bounded and $\left|\hat\psi(u)\right| = \cO_{u\to\infty}(u^{-\eta})$ with $\eta>2$. Let $\varphi_1$ and $\varphi_2$ be bounded functions defined on $\RR^+$ by $\varphi_1(u) = u\left|\hat\psi(u)\right|^2 $ and $\varphi_2(u) = u^2\left|\hat\psi(u)\right|$. Assume $\ccS_X$ is such that 
\[
J_X = \int_0^\infty \xi^{-1}\ccS_X(\xi)\dd\xi < \infty.
\]
Let $\ccS_X^{(k)}$ denote the estimation of the spectrum after $k$ iterations of the algorithm. Let $b_{\ccS_X}^{(k)}$ denote the bias defined for all $m\in[\![ 1,M_s]\!]$ by
\[
b_{\ccS_X}^{(k)}(m) = \EE\left\lbrace\tilde\ccS_X^{(k)}(q^{-s_m}\omega_0)\right\rbrace- \ccS_{X,\psi}(q^{-s_m}\omega_0)\ .
\]
Assume there exists a constant $c_{\theta_1}>0$ such that $\theta_{n,1}^{(k)}>c_{\theta_1}\,,\forall n,k$. Then
\begin{equation}
\left\| b_{\ccS_X}^{(k)} \right\|_{\infty}  \leq\! \dfrac{J_X}{\|\psi\|_{2}^{2}}\left(K'_1\left\|\btheta_1\!-\! \tilde\btheta_1^{(k)}\right\|_\infty\!\!\!+ K'_2\left\|\tilde\btheta_2^{(k)}\!-\!\btheta_2\right\|_{\infty} \right),
\label{fo:spectrum.bias}
\end{equation}
where 
\begin{align*}
K'_1 &= \dfrac{\|\varphi_1\|_\infty}{c_{\theta_1}}<\infty\ ,\\ 
K'_2 &= \ln(q)\left(\|\varphi_1\|_\infty + 2\|\hat\psi'\|_\infty\|\varphi_2\|_\infty\right)<\infty\ .
\end{align*}
\end{proposition}

The proof of the Proposition is given in supplementary materials.
\begin{remark}
If $\btheta_1^{(k)}\to\btheta_1$ and $\btheta_2^{(k)}\to\btheta_2$ as $k\to\infty$, we have $\Ex{\tilde\ccS_X^{(k)}(\nu_m)}\underset{k}\to\ccS_{X,\psi}(\nu_m)$, as expected.
% which is the expected property.

Formula~\eqref{fo:spectrum.bias} enables the control of the spectrum bias at frequencies $\nu_m=q^{-s_m}\omega_0$ only. Notice also that the requirement $J_X<\infty$ forces $\ccS_X$ to vanish at zero frequency.  
%Indeed, the spectrum estimation is obtained thanks to the wavelet transform which is a band-pass filter. Thus zero frequency is not analyzed and when $\ccS_X(0)=0$ none of the information is lost by this filtering. 
\end{remark}

\medskip
\textit{(b)\ Variance.}
The Cram\'er-Rao lower bound (CRLB) gives the minimum variance that can be attained by unbiased estimators. The Slepian-Bangs formula (see~\cite{Stoica97introduction}) directly gives the following CRLB for component $\theta_{n,i}$
$$
\mathrm{CRLB}(\theta_{n,i}) = 2\left(\mbox{Trace}\left\lbrace\left(\bC(\Theta_n)^{-1}\dfrac{\partial\bC(\Theta_n)}{\partial\theta_{n,i}}\right)^2\right\rbrace\right)^{-1}.
$$
This bound gives information about the variance of the estimator at convergence of the algorithm, \ie when both $\ccS_X$ and the other parameters are well estimated.

Applying this formula to $\theta_{n,1}$ gives
$$
\Ex{\left(\tilde\theta_{n,1} - \Ex{\tilde\theta_{n,1}}\right)^2} \geq \mathrm{CRLB}(\theta_{n,1}) = \dfrac{2\theta_{n,1}^2}{M_s}\ .
$$
This implies that the number of scales $M_s$ of the wavelet transform must be large enough to yield an estimator with sufficiently small variance.

For $\theta_{n,2}$, no closed-form expression is available for the CRLB. Therefore, the evaluation of this bound and its comparison with the variance of the estimator $\tilde\theta_{n,2}$ can only be based on numerical results, see section~\ref{se:results}.

\medskip
\textit{(c)\ Robustness to noise.}
Assume now observations are corrupted by a random Gaussian white noise $W$ with variance $\sigma_W^2$ (supposed to be known):
%. The model becomes
\begin{equation}
\label{eq:noisy.model}
Y = \ccA_a\ccD_\gamma X + W\ .
\end{equation}

The estimator $\tilde\theta_{n,1}$ is not robust to noise. Indeed, if the maximum likelihood estimator of model~\eqref{eq:model} in the presence of such white noise, a new term $b_{n,1|_W}^{(k)}\left(\theta_{n,1}\right)$ must be added to the bias expression~\eqref{fo:bias.am}, which becomes
\[
b_{n,1|_W}^{(k)}\left(\theta_{n,1}\right)=\dfrac{1}{M_s}\mbox{Trace}\left\lbrace \tilde\bC_0^{(k)}\left(\tilde\theta_{n,2}^{(k)}\right)^{-1}\bC_{wn}\right\rbrace\,,
\]
where $(\bC_{wn})_{ij} = \sigma_W^2 q^{(s_i+s_j)/2}\int_0^\infty \overline{\hat{\psi}}(q^{s_i}\xi)\hat{\psi}(q^{s_j}\xi)d\xi$. In practice, this term can take large values, therefore noise has to be taken into account. To do so, the covariance matrix is now written as
\begin{equation}
\label{fo:covariance.noise}
\bC(\Theta_n\!)_{ij}\!=\!q^{\frac{s_i+s_j}{2}}\!\!\int_0^\infty\!\!\! (\theta_{n,2}\ccS_{\!X}(q^{-\theta_{n,1}}\xi)+\sigma_{\!W}^2)\overline{\hat{\psi}}(q^{s_i}\xi)\hat{\psi}(q^{s_j}\xi)\!\dd\xi\,
\end{equation}
and the likelihood is modified accordingly. Formula~\eqref{fo:am.estim} is no longer true and no closed-form expression can be derived anymore, the maximum likelihood estimate $\tilde\theta_{n,1}$ must be computed by a numerical scheme (here we use a simple gradient ascent).

The estimator $\tilde\theta_{n,2}$ is very robust to noise. Indeed, equation~\eqref{fo:covariance.noise} shows that the only change in the covariance matrix formula is to replace the power spectrum $\ccS_X$ by $\ccS_Z = \ccS_X + \frac{\sigma_W^2}{\theta_{n,2}}$. The additive constant term does not impair the estimator as long as it is small in comparison with the maximum values of $\ccS_X$.

Moreover, the estimator $\tilde{\ccS_X}$ is modified because when computing $\frac{1}{\theta_{n,1}^{1/2}}\tcW_y\left(s-\theta_{n,2},\tau_n\right)$ on scale $s_m$, we compute:
$$
\bw_{z,s_m} = \bw_{x,s_m} + \bw_{w_*,s_m}\,,
$$
where $ \bw_{w_*,s_m}=\frac{1}{\btheta_{1}^{1/2}}\tcW_w\left(s_m-\btheta_{2},\btau\right)$ is the wavelet transform of a white noise modulated in amplitude by $a^{-1}$.
Thus a constant term $\tilde\sigma_{W}$ independent of frequency is added to the new spectrum estimator $\tilde\ccS_Z$, so that
\begin{equation*}
\Ex{\tilde\ccS_Z} = \ccS_{X,\psi} + \tilde\sigma_W^2\quad \mbox{where}\quad \tilde\sigma_W^2= \sigma_W^2\dfrac{1}{N_\tau}\sum_{n=1}^{N_\tau}\dfrac{1}{\theta_{n,1}}\ .
\end{equation*}

\subsection{Extension: estimation of other deformations}
To describe other nonstationary behaviors of audio signals, other  operators can be investigated. For example, combination of time warping and frequency modulation can be considered, as was done in~\cite{Meynard17spectral}, we shortly account for this case here for the sake of completeness. Let $\alpha\in C^2$ be a smooth function, and set
\beq
\label{fo:freq.modul}
\ccM_\alpha:\qquad \ccM_\alpha x(t) = e^{2\ii\pi\alpha(t)}x(t)\ ,
\ee
The deformation model in~\cite{Meynard17spectral} is of the form
\begin{equation}
\label{eq:modul.model}
Y = \ccA_a\ccM_\alpha\ccD_\gamma X\ .
\end{equation}
To perform joint estimation of amplitude and frequency modulation and time warping for each time, a suitable time-scale-frequency transform $\cV$ is introduced, defined as $\cV_X(s,\nu,\tau) = \langle X,\psi_{s\nu\tau} \rangle$, with $\psi_{s\nu\tau} = T_\tau M_\nu D_s\psi$. In that case, approximation results similar to Theorem ~\ref{th:approx} can be obtained from which the corresponding log-likelihood can be written.
At fixed time $\tau$, the estimation strategy is the same as before, but the parameter space is of higher dimension, and the extra parameter  $\theta_3=\alpha'(\tau)$ complicates the log-likelihood maximization. In particular, the choice of the discretization of the two scale and frequency variables $s$ and $\nu$ influences performances of the estimator, in particular the Cram\'er-Rao bound.

\section{Numerical results}
\label{se:results}

We now turn to numerical simulations and applications. A main ingredient is the choice of the wavelet transform. Here we shall always use the \textit{sharp wavelet} $\psi_\sharp$ defined in~\eqref{fo:wavelet} and set the scale constant to $q=2$.
%The choice of $\epsilon$ also enables to fix the bandwidth of the wavelet. Indeed, denoting the 3~dB bandwidth by $\Delta_\omega$, we have $\frac{\Delta_\omega}{\omega_0} = \sqrt{C_\epsilon(C_\epsilon+4)}$ where $C_\epsilon = \frac{-2\ln(2)}{\ln(\epsilon)}\delta(\omega_1,\omega_0)$.

We systematically compare our approach to simple estimators for amplitude modulation and time warping, commonly used in applications, defined below. The approach of~\cite{Clerc03estimating} was also implemented, but we couldn't get satisfactory results with that approach.
\begin{itemize}
\item
Amplitude modulation: we use as baseline estimator of $a(\tau_n)^2$ the average energy $\tilde\theta_{n,1}^{(B)}$  defined as follows:
\[
\tilde\theta_{n,1}^{(B)} = \dfrac{1}{M_s}\|\bw_{y,\tau_n}\|^2\,.
\]
This amounts to replace the estimated covariance matrix in~\eqref{fo:am.estim} by the identity matrix. Notice that $\tilde\theta_{n,1}^{(B)}$ does not depend on the time warping estimator, and can be computed directly on the observation.
\item
Time warping: the baseline estimator $\tilde\theta_{n,2}^{(B)}$ is the scalogram scale center of mass defined as follows:
\[
\tilde\theta_{n,2}^{(B)} = C_0 + \dfrac{1}{\|\bw_{y,\tau_n}\|^2}\sum_{m=1}^{M_s}s[m]|\bw_{y,\tau_n}[m]|^2\,.
\]
$C_0$ is chosen such that $\tilde\btheta_{2}^{(B)}$ is a zero-mean vector. 
\end{itemize}
Numerical evaluation is performed on both synthetic signals and deformations and real audio signals.

\subsection{Synthetic signal}
We first evaluate the performances of the algorithm on a synthetic signal. This allows us to compare variance and bias with their theoretical values.

The simulated signal has length $N_\tau=2^{16}$ samples, sampled at $F_s = 8$~kHz  (meaning the signal duration is $t_F=(N_\tau-1)/F_s\approx 8.2$~s). The spectrum $\ccS_X$ is written as $\ccS_X = S_1 + S_2$ where $S_l(\nu) = 1+\cos\left(2\pi(\nu-\nu_0^{(l)})/\Delta_\nu^{(l)}\right)$ if $|\nu-\nu_0^{(l)}|<\Delta_\nu^{(l)}/2$ and vanishes elsewhere (for $l\in\lbrace 1,2\rbrace$). The amplitude modulation $a$ is a sine wave $a(t) = a_0\left(1 + a_1\cos(2\pi t/T_1)\right)$, where $a_0$ is chosen such that $t_F^{-1}\int_0^{t_F}a^2(t)dt=1$. The time warping function $\gamma$ is  such that $\log_q(\gamma'(t)) = \Gamma + \cos(2\pi t/T_2)e^{-t/T_3}$, where $\Gamma$ is chosen such that $t_F^{-1}\int_0^{t_F}\gamma'(t)dt=1$.

\begin{figure}
  \centering
  \includegraphics[width=0.7\textwidth]{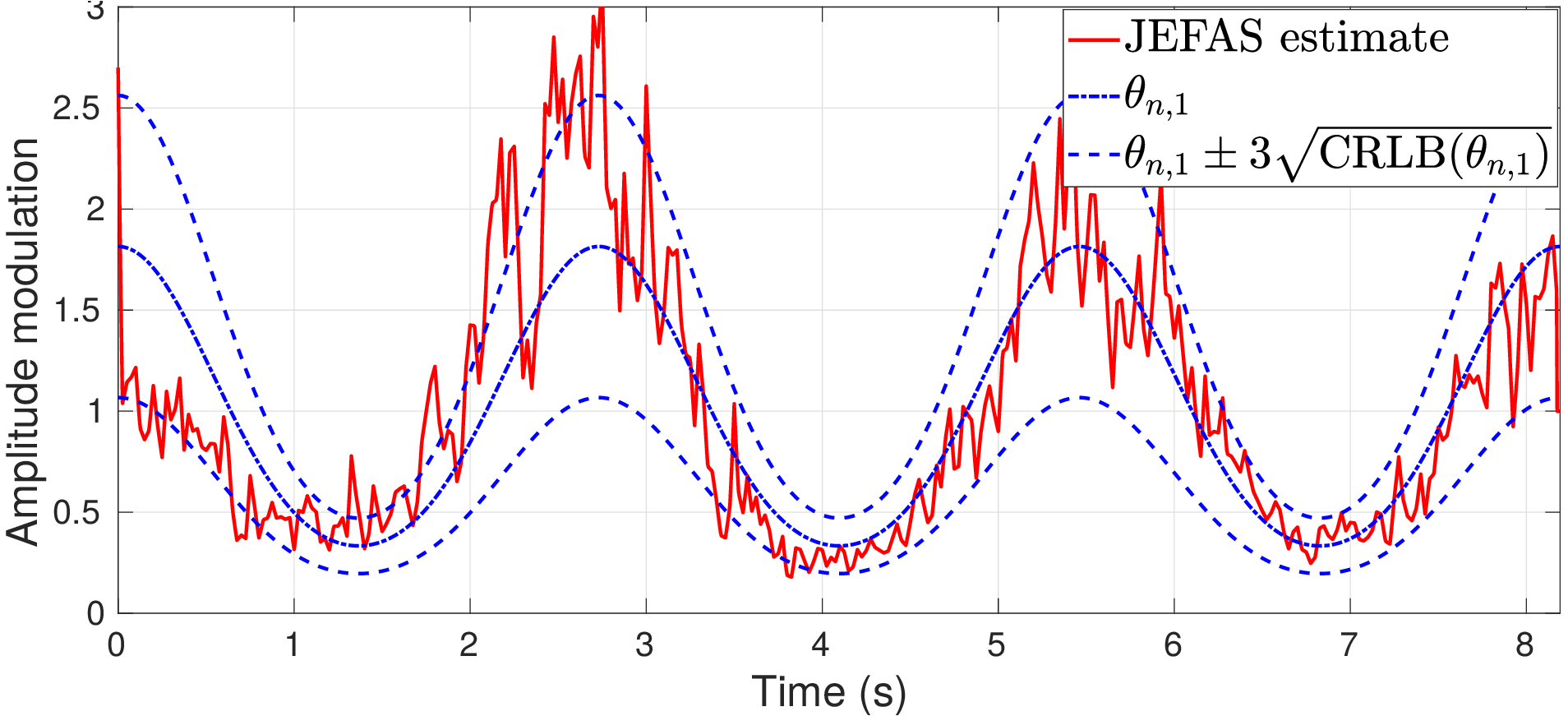}\\ % single column
  \includegraphics[width=0.7\textwidth]{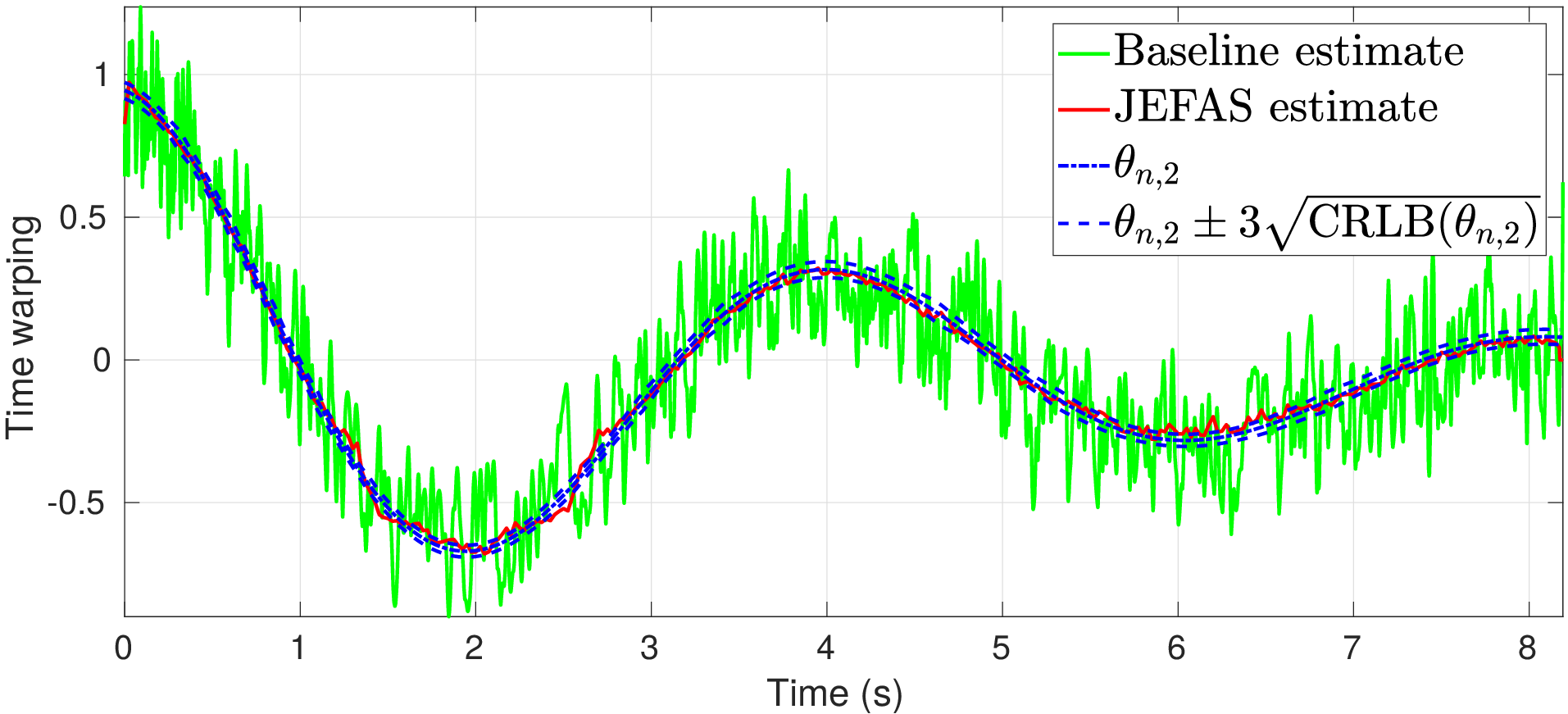}% single column
  \caption{Joint amplitude modulation/time warping estimation on a synthetic signal. Top: amplitude modulation estimation ($a_1=0.4$ and $T_1=t_F/3$). Bottom: warping estimation ($T_2=t_F/2$ and $T_3=t_F/2$).}
\label{fig:deform.estimations}
\end{figure}

JEFAS is implemented in the MATLAB/Octave scientific environment. Dimensions were set as $M_s = 106$ and $p = 7$. The wavelet transform is computed using the sharp wavelet with $\ln(\epsilon)=-25$ corresponding to a quality factor $Q=6$. In this problem, the algorithm took 67 seconds to converge on a standard desktop computer (CPU Intel Core @ 3.20~GHz $\times$ 4, 7.7 GB RAM). Results are shown in Fig.~\ref{fig:deform.estimations} and compared with baseline estimations. For the sake of visibility, the baseline estimator of the amplitude modulation (which is very oscillatory) is not displayed, but numerical assessments are provided in Table~\ref{tab:mse}, which gives MSEs for the different estimations. JEFAS is clearly more precise than the baseline algorithm, furthermore its precision is well accounted for by the Cram\'er-Rao bound: in Fig.~\ref{fig:deform.estimations}, the estimate is essentially contained within the 95~\% confidence interval provided by the CRLB (assuming Gaussianity and unbiasedness).

\begin{table}
\centering
\begin{tabular}{|l||c|c|}
  \hline
  Estimation & Amplitude  & Time  \\
  method &  modulation  & warping \\
  \hline
  \hline
  Baseline & $0.2015$ & $0.0232$ \\
  \hline
  JEFAS & $0.0701$ & $0.0005$ \\
  \hline
\end{tabular}
\medskip
\caption{Estimation mean square errors for both deformations}
\label{tab:mse}
\end{table}

\begin{figure}
\begin{center}
  \includegraphics[width=0.4\textwidth]{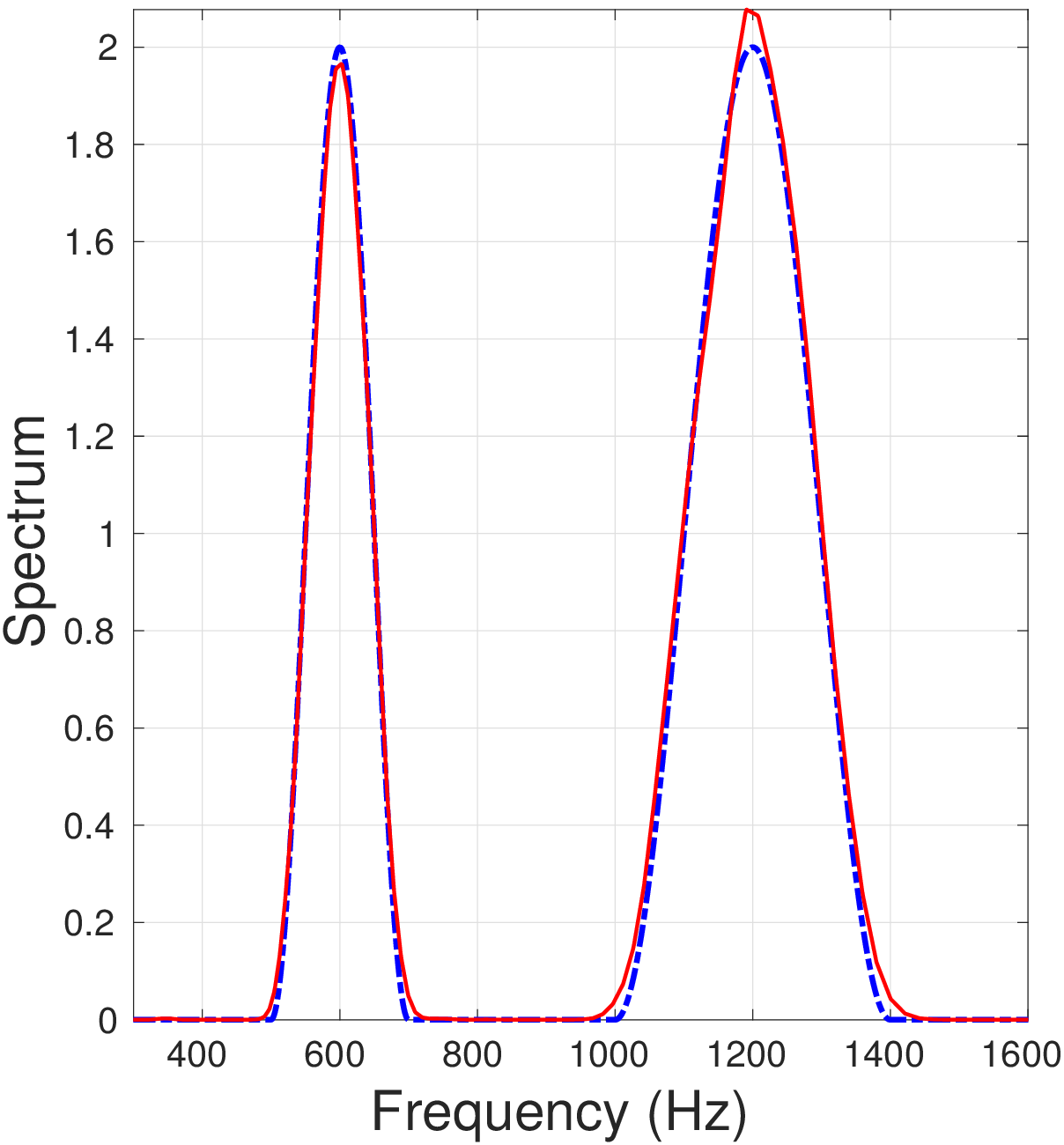}% single column
  \includegraphics[width=0.4\textwidth]{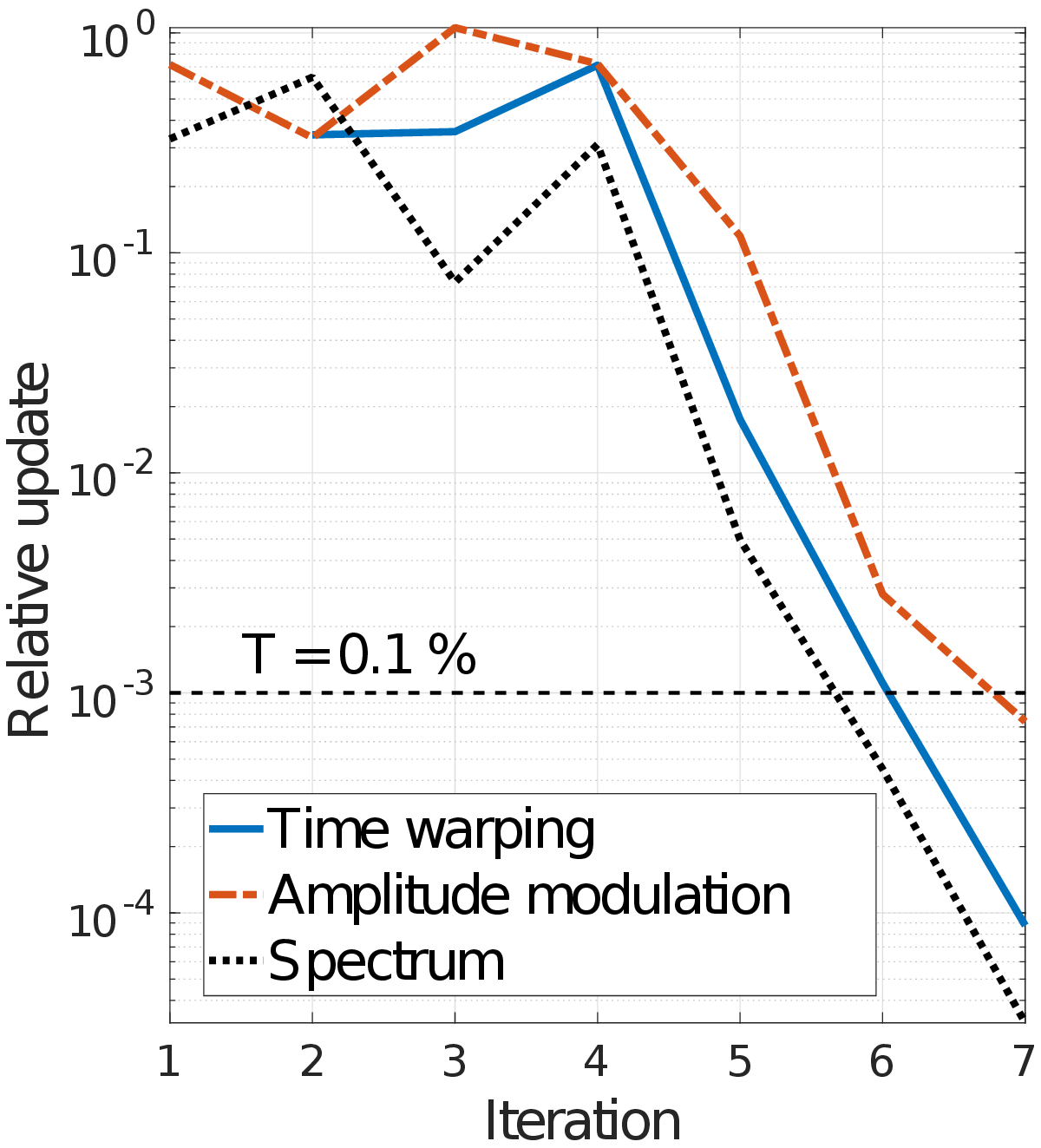}% single column
\end{center}
  \caption{Left: Spectrum estimation ($\nu_0^{(1)}=600$~Hz, $\Delta_\nu^{(1)} = 200$~Hz, $\nu_0^{(2)}=1.2$~kHz, $\Delta_\nu^{(2)} = 400$~Hz): actual (dash-dot blue line) and estimated (solid red line) spectra. Right: Relative update evolution.}
\label{fig:spectrum.estimation}
\end{figure}

The left hand side of Fig.~\ref{fig:spectrum.estimation} displays the estimated spectrum given by formula~\eqref{fo:spectrum.estimator}. The agreement with the actual spectrum is very good, with a slight enlargement effect due to filtering by $|\hat{\psi}|^2$.
The right hand side of Fig.~\ref{fig:spectrum.estimation} gives the evolution of the stopping criterion~\eqref{fo:stop.crit} with iterations. Numerical results show that time warping estimation converges faster than amplitude modulation estimation. Nevertheless, when fixing a stopping criterion to 0.1~\% only 7 iterations are necessary for JEFAS to converge.

\subsection{Application to dolphin sound spectral analysis}
\begin{figure}
\begin{center}
  \includegraphics[width=0.9\textwidth]{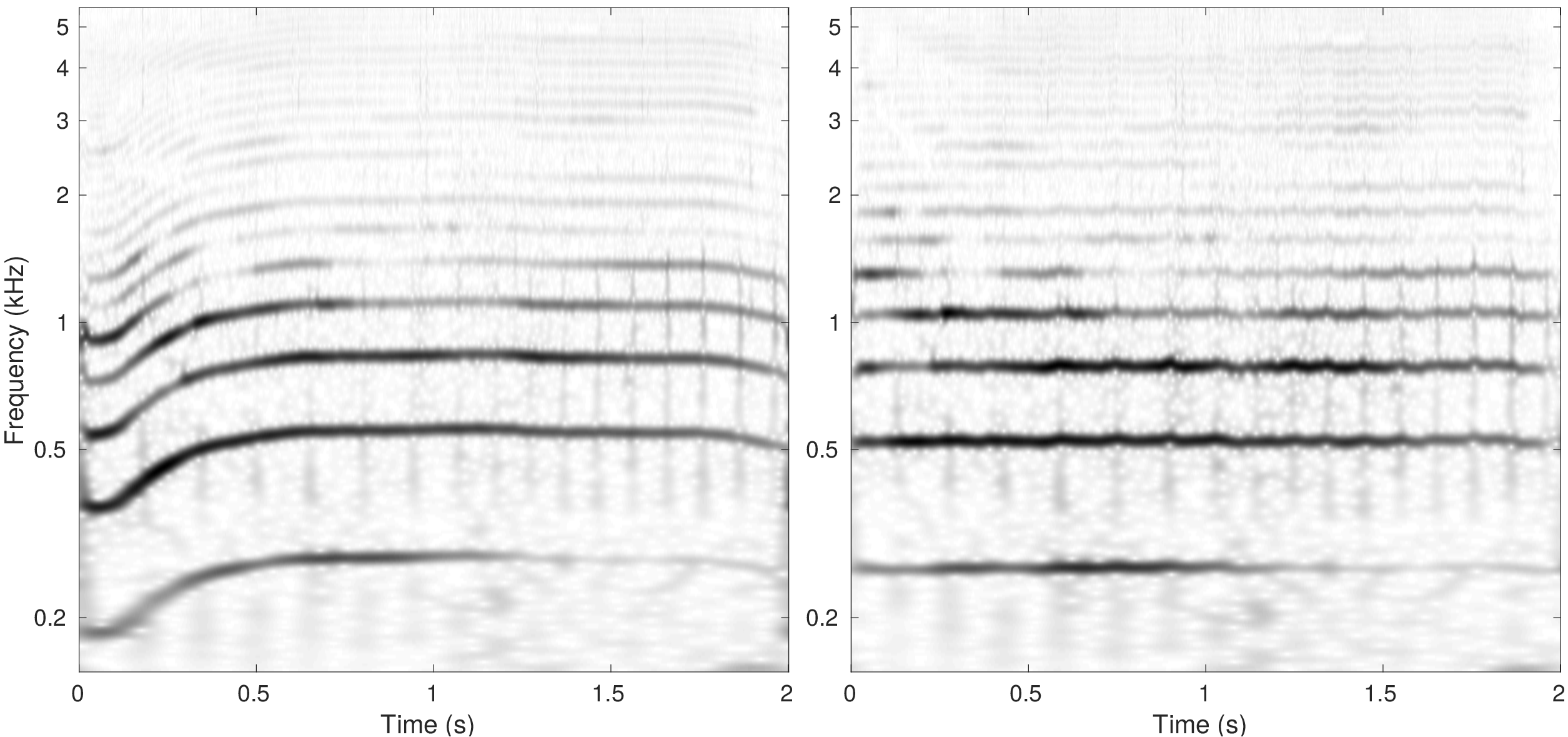} 
  \includegraphics[width=0.9\textwidth]{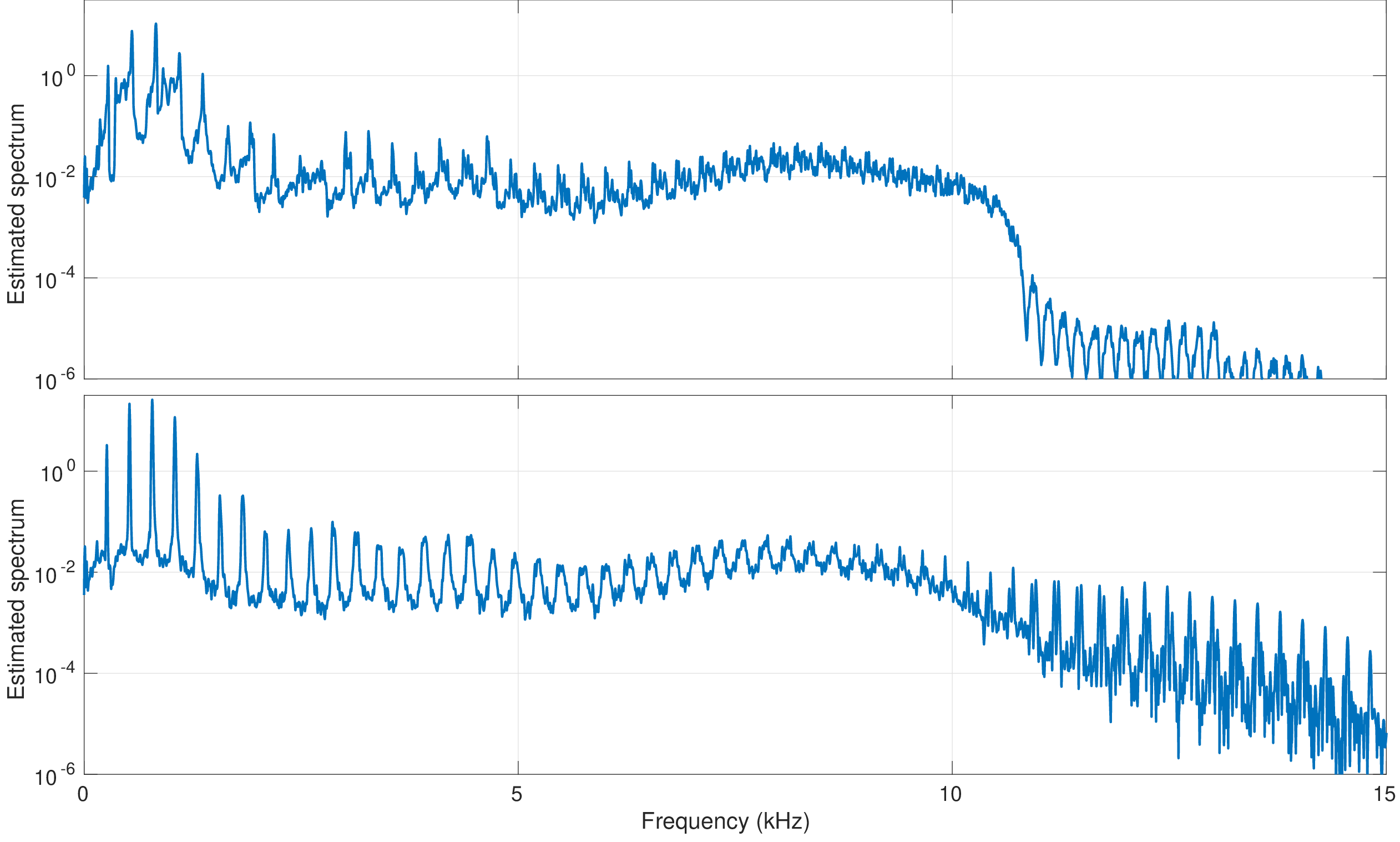}
\end{center}
  \caption{Dolphin sound spectral analysis. Top left: log-scalogram of the original signal. Top right: log-scalograms of unwarped and unmodulated signals. Middle: estimated spectrum from the original signal. Bottom:  spectrum from the estimated underlying stationary signal.}
\label{fig:dolphin}
%\vspace{-2mm}
\end{figure}

After studying the influence of the various parameters, we now turn to real-world audio examples.
First, we analyze a recording of a two seconds long dolphin vocalization sound, described in~\cite{Stowell18computational}. The wavelet transform of this signal in Fig.~\ref{fig:dolphin} shows that the warping model~\eqref{eq:model} fits well this kind of signal, except for transient clicks that are not accounted for. JEFAS allows the estimation of the spectrum of the underlying stationary signal.

On the top-right of Fig.~\ref{fig:dolphin}, we display the wavelet transform of the signal obtained by application of the inverse deformations estimated by JEFAS. Notice that the presence of clicks slightly disturbs the stationarization process. Nonetheless, it makes sense to estimate a power spectrum from this signal, since the time dependence of its wavelet transform is negligible with respect to its scale dependence. The estimated spectra from the original signal and from the estimated underlying stationary signal are displayed on the middle and the bottom of Fig.~\ref{fig:dolphin}. Thanks to JEFAS, the harmonic structure clearly appears (bottom plot). We believe the application of JEFAS to these types of sounds can potentially bring new insights in bioacoustic applications.

\subsection{Application to Doppler estimation}
Finally, we analyze a sound which is a recording (from a fixed location) from a racing car, moving with constant speed. The car engine sound is then deformed by the Doppler effect, which results in time warping, as explained below. Besides, as the car is moving, the closer the car to the microphone, the larger the amplitude of the recorded sound. Thus, our model fits well this signal.

\begin{figure}
\begin{center}
  \includegraphics[width=0.9\textwidth]{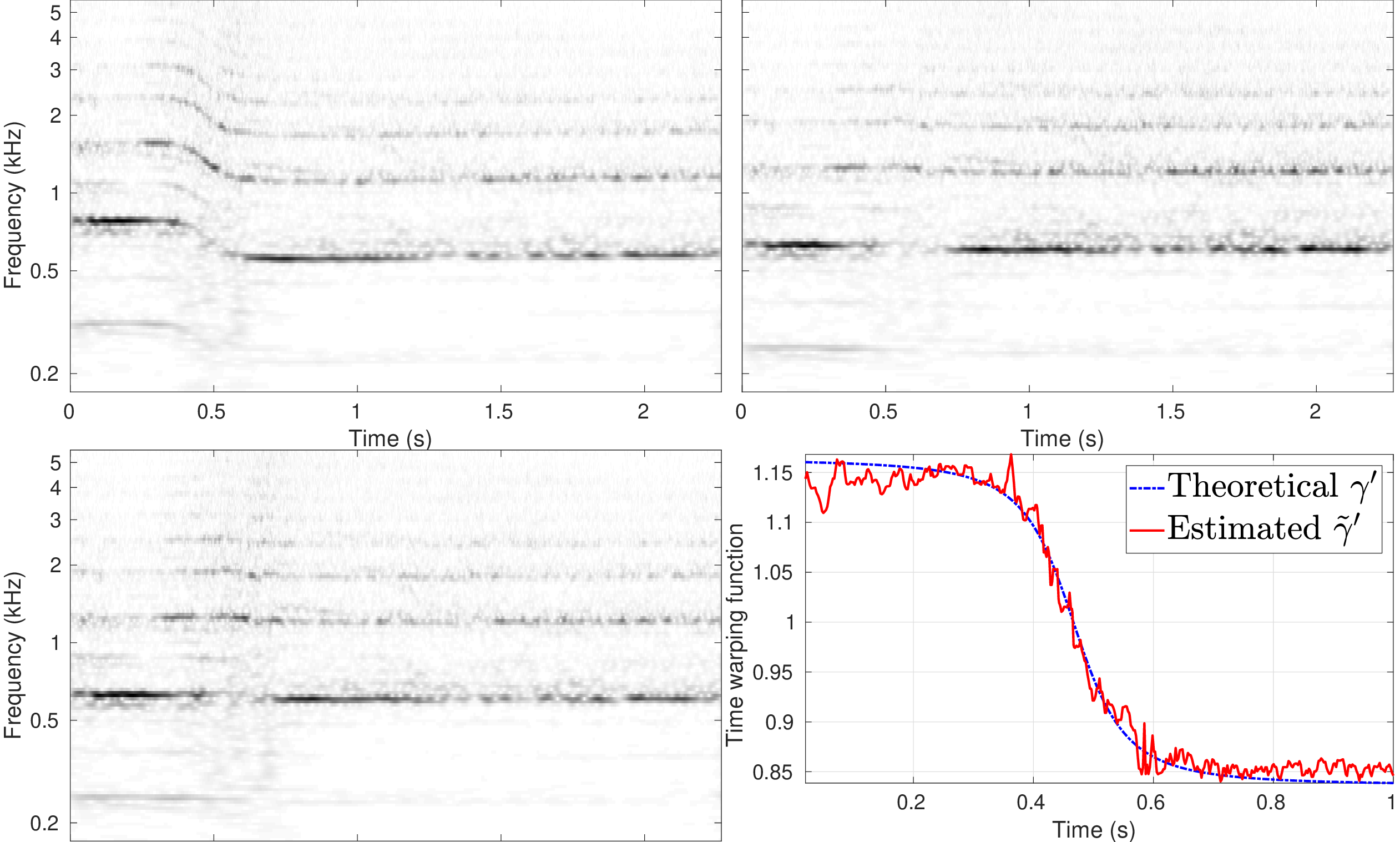}% single column
\end{center}
  \caption{Doppler estimation. Top:  log-scalograms of the original (left) and unwarped and unmodulated (right) signals. Bottom left: log-scalogram of the unwarped signal. Bottom right: Estimated time warping compared with the theoretical value given in~\eqref{fo:doppler}.}
\label{fig:f1}
%\vspace{-4mm}
\end{figure}

The wavelet transforms of the original signal and the two estimations of the underlying stationary signal are shown in Fig.~\ref{fig:f1}. While the estimation of time warping only corrects the displacement of wavelet coefficients in the time-scale domain, the joint estimation of time warping and amplitude modulation also approximately corrects nonstationary variations of the amplitudes.
  
The physical relevance of the estimated time warping function can be verified. Indeed, denote by $V$ the (constant) speed of the car and by $c$ the sound velocity. Fixing the time origin to the time at which the car passes in front of the observer at distance $d$, the time warping function due to Doppler effect can be shown to be
\begin{equation}
\gamma'(t) = \dfrac{c^2}{c^2-V^{2}}\left( 1 - \dfrac{V^2t}{\sqrt{d^2(c^2-V^2) + (cVt)^2}} \right) _ .
\label{fo:doppler}
\end{equation}
We plot in Fig.~\ref{fig:f1} (bottom right) the estimation $\tilde\gamma'$ compared with its theoretical value where $d = 5$~m and $V=54$~m/s. Clearly the estimate is close to the corresponding theoretical curve obtained with these data, which are therefore realistic values.

Nevertheless, a closer look at scalograms in Fig.~\ref{fig:f1} shows that the amplitude correction is still not perfect, due to the presence of noise, and the fact that the model remains too simple: the amplitude modulation actually depends on frequency, which is not accounted for.

%\subsection{Time warping and frequency modulation}
%
%{\color{red}
%- Pas de signaux r\'eels correspondant \`a ce mod\`ele $=>$ Pas de r\'esultats
%
%- Justifier du fait qu'une estimation convenable est impossible car la borne de Cramer-Rao est grande. 
%}

\section{Conclusions}
We have discussed in this paper extensions of methods and algorithms described earlier in~\cite{Omer13estimation,Omer15Modeles,Omer17time,Meynard17spectral} for the joint estimation of deformation operator and power spectrum for deformed stationary signals, a problem already addressed in~\cite{Clerc03estimating} with a different approach. Besides some improvements on the estimation algorithm itself, the main improvements described in this paper concern the following two points
\begin{enumerate}
\item
the extension of the algorithm to the joint estimation of deformations including amplitude modulation to the model and its estimation (\cite{Meynard17spectral} was limited to time warping and combinations of time warping with frequency modulation, and investigated generalized wavelet transforms);
\item
a statistical study of the estimators and of the performances of JEFAS algorithm, with precise mathematical statements.
\end{enumerate}
The proposed approach was validated on numerical simulations and applications to two case studies: spectral estimation from non-stationary dolphin vocalization, and Doppler estimation.

The results presented here show that the proposed extensions yield a significant improvement in terms of precision, and a better theoretical control. In particular, the continuous parameter estimation procedure avoids quantization effects that were present in~\cite{Omer17time} where the parameter space was discrete and the estimation based on exhaustive search. It also allows the derivation of precision estimates, in particular a Cram\'er-Rao bound. Numerical results show that the introduction of amplitude modulation also improves results. Finally, regarding the approach of~\cite{Clerc03estimating}, its domain of validity seems to be limited to small-scale (\ie high-frequency) signals, which is not the case here.

Contrary to~\cite{Clerc03estimating}, our approach is based on (approximate) maximum likelihood estimation in the Gaussian framework. Because of our choice to disregard time correlations, the estimates obtained here generally present spurious fluctuations, which can be smoothed out by appropriate filtering. A natural extension of our approach would be to introduce a smoothness prior that should avoid such filtering steps when necessary.

We believe that being able to estimate precisely warping functions can be valuable in a variety of audio applications. Speech applications have already been studied, we may also mention bioacoustic signals, for example to refine the frequency excursion indices used to assess vocal performances of songbirds (see~\cite{Podos16finescale} and references therein). Quite obviously, controlling warping and spectrum opens new perspectives in sound design, for example for cross-synthesis. Future work will apply the aforementioned methods to a task of blind source separation of nonstationary signals.

\smallskip
The code and datasets used to produce the numerical results of this paper, and other audio examples (female voice, wind, etc.) are available at the web site
\begin{center}
\small\tt
https://github.com/AdMeynard/JEFAS
%https://hal.archives-ouvertes.fr/hal-01670187
\end{center}
More details on the sharp wavelet and proofs of Theorem~\ref{th:approx} and Proposition~\ref{pr:spectrum.bias} are given as supplementary material, that also includes another case study (application to wind sound).

%\appendices
%\input{appendices}
%\vspace{-1mm}
\section*{Acknowledgment}
We wish to thank M. Kowalski and R. Kronland-Martinet for fruitful discussions. We also thank the anonymous reviewers for many valuable comments and suggestions, and for bringing very interesting references to our attention. We also thank the Centre de Recherches Math\'ematiques of Universit\'e de Montr\'eal, where part of this research has been done, for kind hospitality.

%\vspace{-1mm}
\bibliographystyle{IEEEtran}
\bibliography{Sampta17}

%\begin{IEEEbiography}{Adrien Meynard}
%Biography text here.
%\end{IEEEbiography}
%\begin{IEEEbiography}{Bruno Torr\'esani}
%Biography text here.
%\end{IEEEbiography}

\end{document}